\documentclass[aps,pra,twocolumn,showpacs,preprintnumbers,amsmath,amssymb,floatfix,superscriptaddress]{revtex4-1}

\usepackage{ulem}

\usepackage{graphics}
\usepackage{graphicx}
\usepackage{amsmath}
\usepackage{setspace}
\usepackage{braket}
\usepackage{epstopdf}
\usepackage{comment}
\usepackage{hyperref}
\usepackage{bm}
\usepackage{xfrac}
\usepackage{blindtext}
\usepackage{mathrsfs}
\usepackage{xcolor}
\usepackage[version=4]{mhchem}

\usepackage{chngcntr} 

\newcommand{\affA}{Fritz-Haber-Institut der Max-Planck-Gesellschaft, Faradayweg 4-6, 14195 Berlin, Germany }
\newcommand{\affB}{Department of Physics and Astronomy, Stony Brook University, Stony Brook, New York 11794, USA}

\newcommand{\affC}{Institute for Advanced Computational Science, Stony Brook University, Stony Brook, NY 11794-3800, USA}

\begin{document}

\title{The chemistry of AlF and CaF production in buffer gas sources}

\author{X.~Liu}\affiliation{\affA}
\author{W.~Wang}\affiliation{\affA}
\author{S. C.~Wright}\affiliation{\affA}
\author{M.~Doppelbauer}\affiliation{\affA}
\author{G.~Meijer}\affiliation{\affA}
\author{S.~Truppe}\affiliation{\affA}
\author{J.~P\'erez-R\'{i}os}\affiliation{\affB}\affiliation{\affC}\affiliation{\affA}

\date{\today}

\begin{abstract}

In this work, we explore the role of chemical reactions on the properties of buffer gas cooled molecular beams. In particular, we focus on scenarios relevant to the formation of AlF and CaF via chemical reactions between the Ca and Al atoms ablated from a solid target in an atmosphere of a fluorine-containing gas, in this case, \ce{SF6} and \ce{NF3}. Reactions are studied following an \textit{ab initio} molecular dynamics approach, and the results are rationalized following a tree-shaped reaction model based on Bayesian inference. We find that \ce{NF3} reacts more efficiently with hot metal atoms to form monofluoride molecules than \ce{SF6}. In addition, when using \ce{NF3}, the reaction products have lower kinetic energy, requiring fewer collisions to thermalize with the cryogenic helium. Furthermore, we find that the reaction probability for AlF formation is much higher than for CaF across a broad range of temperatures.

\end{abstract}

\maketitle



\section{Introduction}

All routes to ultracold molecules require and produce molecules in a small and well-defined number of internal states. Experiments generally separate into those using indirect or direct cooling techniques. Indirect cooling techniques refer to scenarios in which ultracold atoms are bound together to form an ultracold molecule via photoassociation~\cite{Lett1995,PA,PA2,PAJPR} or magnetoassociation~\cite{Moerdijk,Magnetoassociation}, which is customary for ultracold bi-alkali molecules. Direct cooling techniques involve producing molecules in a source and extracting them into a molecular beam that is slowed and captured in a trap. The manipulation techniques involved include Stark deceleration~\cite{Stark}, Zeeman slowing~\cite{Zeeman}, centrifuge deceleration~\cite{Cherenkov} and laser and optoelectrical cooling~\cite{lasercooling,Prehn2016}. Similarly, heat dissipation can be obtained by bringing an ensemble of molecules in contact with a cold reservoir, typically He atoms, known as buffer gas cooling~\cite{Weinstein,Buffergas1,Buffergas2,Hutzler2021}, or sympathetic cooling, in which direct cooling of an ensemble of one species facilitates cooling of another~\cite{sympa1,sympa2,Tobias}.

Production of diatomic molecules in a cryogenic buffer gas is a versatile method to create an ensemble of molecules in its lowest internal states and has been applied to many molecular species. The thus produced cold molecules can be directly trapped in a magnetic field~\cite{Weinstein}, for instance, or cold molecules emanating from the buffer gas source in an effusive beam can be used for further cooling, deceleration, and trapping experiments. In many of these buffer gas sources, molecules are produced in a chemical reaction of laser-ablated atoms — that have an initial temperature of several thousand Kelvin — with a reactant gas. The molecules subsequently thermalize with the cryogenic buffer gas. This process spans several orders of magnitude in energy, involves complex reaction kinetics, and is thus far not well understood. Indeed, only recently, molecular beams emerging from a buffer gas source have been successfully simulated based on general physical properties~\cite{Schullian2015, Gantner2020,Hutzler2021}. However, to the best of our knowledge, the impact of the reactive gas where ablation occurs in buffer gas sources remains unexplored, both experimentally and theoretically. Furthermore, properties of the molecular beam such as its overall yield, short- and long-term stability, and phase-space distribution ultimately determine which downstream experiments are possible. Therefore, it is mandatory to understand the chemistry in buffer gas cells to design brighter and colder molecular beams facilitating the application of subsequent cooling techniques.

This paper presents a first principle study to understand the role of chemical reactions in a cryogenic buffer gas source. As an example, we focus on the formation efficiency of AlF/CaF after laser-ablating Al/Ca in two distinct fluorine-donor gases: \ce{NF3} and \ce{SF6}. Recently, we have demonstrated that a molecular beam of \ce{AlF} \cite{Hofsaess2021} is about an order of magnitude brighter as compared to a beam of CaF \cite{Truppe2017}. The aim is to understand the reaction in more detail and analyze the impact of the specific fluorine donor gas on the molecular beam properties. After performing molecular dynamics simulations, we demonstrate that the number and kinetic energy of the product molecules depends on the nature of the fluorine donor gas as much as on the intensity of the ablating laser, which sets the reaction temperature. Tree-shaped reaction models are constructed to understand further the impact of stereochemistry in these reactions to reveal critical intermediate states and their transitions in the reaction channels. These results establish the basis for a more detailed understanding of the properties of buffer gas molecular beams, which will aid in their future design and optimization.

\section{Computational details and methodology}

In a buffer gas cell, highly energetic ablated metal atoms collide with lower energy fluorine-containing molecules present. This scenario defines a non-equilibrium system. Therefore, it requires formulating the problem in the grand canonical ensemble, which is theoretically cumbersome and computationally extremely expensive. Nevertheless, since the temperature of the ablated atoms is much higher than the buffer gas temperature, the atom temperature is a good approximation for the reaction temperature~\footnote{This is strictly true before the time it takes the ablated atoms and reactant molecule to collide like 10-20 times with buffer gas atoms. Indeed, it is fulfilled here since the reaction time is $\sim 1$~ps, whereas the collision time for He-F-containing molecule is 500~ns assuming a He-molecule elastic cross section $\sim 10^{-14}$~cm$^{-2}$ and He density of 10$^{21}$m$^{-3}$.}. In this scenario, we launch $N_t$ trajectories according to a given temperature via the Maxwell-Boltzmann distribution. The time evolution of each of these trajectories is computed via {\it ab initio} molecular dynamics (AIMD) to elucidate the reaction dynamics for Al/Ca+\ce{NF3}/\ce{SF6} reactions within the microcanonical ensemble --also known as the NVE ensemble since the number of particles, volume and energy are conserved quantities-- using the hybrid BHLYP functional~\cite{becke1988density,becke1993new}, previously reported to reproduce the experimental rate constants for Al + \ce{SF6} reaction~\cite{parker2002kinetics}. In this work, we also include the D3 dispersion correction~\cite{grimme2010consistent} to reach a better description of the ubiquitous long-range interactions. All these calculations are performed employing the def2-TZVP basis set~\cite{basiskaupp1991pseudopotential,basisleininger1996accuracy,basisweigend2005balanced}, as implemented in the Gaussian 16 package~\cite{g16}.

\subsection{Initial conditions}

\begin{figure}[t]
    \centering
    \includegraphics[width=0.7\linewidth]{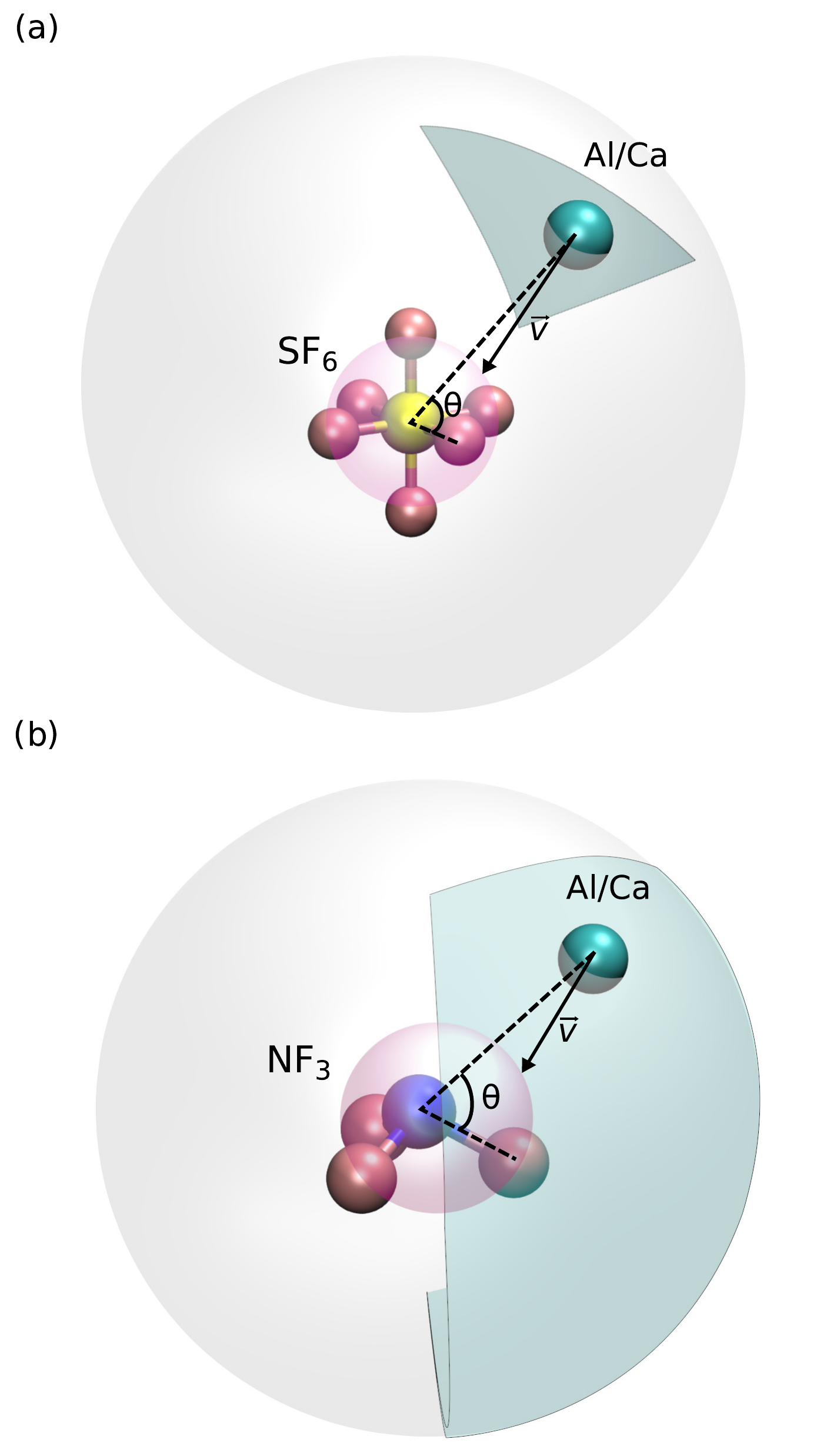}
    \caption{Initial positions of Al/Ca atoms in the AIMD simulations, randomly sampled from a symmetry-reduced sphere of radius $7$ \AA, centered at the S or N atoms. The velocity vector direction is randomly sampled to ensure an atom-molecule collision as the pink surface sketches it.}
    \label{Fig:initial_positions}
\end{figure}

For each of the $N_t$ trajectories, the initial position of the metal atom is randomly selected from a sphere of radius 7~\AA, centered at atom (\ce{N/S}) of the target molecule (\ce{NF3}/\ce{SF6}), as it is shown in Fig.\ref{Fig:initial_positions}. We use the molecule's symmetry to reduce the angular degrees of freedom range to improve the sampling efficiency. The magnitude of the metal atoms' velocity satisfies a Maxwell-Boltzmann distribution at corresponding temperatures, and its direction is randomly sampled following the polar and azimuthal angles in spherical coordinates to ensure a collision with the target molecule as sketched by the pink surface in Fig.\ref{Fig:initial_positions}.

\subsection{Reaction probability}
The main descriptor of a chemical process is the reaction probability, defined as the probability that the reactants end up in a particular product state. Mathematically, the reaction probability is defined as 

\begin{equation}
    \mathcal{P}_r=\frac{N_r}{N_t},
\end{equation}
where $N_r$ denotes the number of trajectories leading to a given reaction product. In particular, we take $N_t=1000$ up to a final state in which the metal atom is at a distance larger than $7~\textrm{\AA}$ from the N or S atom for each temperature and colliding species. Each of these trajectories takes, on average, 4.5 hours to run on 40 processors.

\subsection{Reaction model}

Reaction models, providing a microscopic understanding of reaction mechanism, are prerequisites for sensitivity analysis and reaction route design
\cite{reactionmodelautomaticgao2016reaction,reactionmodelgraphsakamoto1988graph,reactionmodelkarstenmargraf2019systematic,reactionmodelontheflybroadbelt1994computer,reactionmodelsimm2017context,reactionmodelkim2018efficient,reactionmodeldeshpande2020graph,reactionmodelulissi2017address,reactionmodelgupta2021learning,reactionmodelrappoport2014complex,reactionmodelmutlay2015complex,reactionmodelzeng2020complex}. To construct such models, generally one considers the thermodynamic properties of the chemical species at the elementary steps in the reaction, including the equilibrium geometries and transition states. In this case, transition probabilities between these species are calculated by the transition state theory. As a result, the mean-field reaction probability of each reaction step can be obtained from Eyring's theory under the equilibrium hypothesis \cite{reactioneyring1935activated}. However, for reactions at finite temperature, like the ones in this work, for which the collision energy of the system is high, the system may go through reaction pathways that deviate from minimal energy paths so that the transition state theory can no longer apply\cite{reactionparrinello2003efficient,reactionsun2015ab,reactionfoppa2019co}.

On the other hand, trajectory-based AIMD simulations provide direct information about reaction pathways, including temperature effects and transition probabilities between the sampled species. Indeed, AIMD-simulated reaction trajectories have been featured in a discrete structural space to construct reaction models \cite{reactionwang2021complex}. In addition, to further understand the impact of temperature and stereochemistry in these reactions, we have developed a tree-shaped reaction model revealing the relevance of intermediate states and transitions in every reaction channel. Such a reaction model relies on identifying the most characteristic states of the reaction and estimating the transition probabilities between each of them via Bayesian inference~\cite{BayesianInference}.

In our reaction model, a molecular structure is represented in terms of the vector $\mathbf{g}$ of atom-pairs-inverse-distances, 
\begin{equation}
    \mathbf{g}_i = [\frac{1}{||\mathbf{m}-\mathbf{n}||}], (\mathbf{m} \neq \mathbf{n}; \mathbf{m}, \mathbf{n} \in \mathbf{r}_i),
\end{equation}
where $\mathbf{r}_i$ represents the Cartesian coordinates of the $i$th MD step. As a result, a trajectory can be expressed as $\mathbf{G} = \{\mathbf{g}_1, \dots, \mathbf{g}_i, \dots, \mathbf{g}_n\}$, with $n$ being the number of MD time steps. Next, using a regular pace clustering algorithm~\cite{regularspaceclustering}, for each reaction studied, we identify 13 states. The clustered trajectory can be represented as $\mathbf{X}=\{\mathbf{x}_1,\dots,\mathbf{x}_i,\dots,\mathbf{x}_n\}$, where $i \in \{0-12\}$.

Calculating the transition probabilities among the 13 states requires a prior distribution, and in this case, we assume a Bernoulli distribution given by the following mass function
\begin{equation}
    f(\delta_{\mathbf{x}};p) = p^{\delta_{\mathbf{x}}}(1-p)^{1-\delta_{\mathbf{x}}},
\end{equation}
where $\delta_{\mathbf{x}}$ is the Kronecker delta function, which equals to $1$ when state $\mathbf{x}$ is observed and $p=0.5$. The probability of a successful transition to each of the states is the same. Then, the transition matrix $\mathbf{T}$ can be defined as
\begin{equation}
    \mathbf{T} = [p_{ij}] \in \Re_{\geqslant 0}
\end{equation}
where $p_{ij}$ is the posterior probability of state $i$ to successfully transfer into state $j$, and it is calculated via Markov chain Monte Carlo (MCMC) implemented in PyMC3\cite{pymc3}. In addition, it is worth mentioning that the present reaction model is essentially a Bayesian network\cite{kjaerulff2008bayesian}, since the posterior probabilities are trained by feeding the clustered trajectories $\{\mathbf{X}_i\}$ ($i \in N_t$).

To analyze in detail the product selectivity, we further consider the states that determine the products, i.e., states $9$ to $12$. The vector $\mathbf{g}'$, which is the inverse distances between \ce{N}/\ce{S} atom and \ce{F} atoms, is calculated as
\begin{equation}
    \mathbf{g}'_i = [\frac{1}{||j-k||}], (j \in \mathbf{r}_i^{\ce{N}}/\mathbf{r}_i^{\ce{S}}; k \in \mathbf{r}_i^{\ce{F}}; \mathbf{x}_i \in \{9-12\})
\end{equation}
where $\mathbf{r}_i^{\ce{N}}$, $\mathbf{r}_i^{\ce{S}}$ and $\mathbf{r}_i^{\ce{F}}$ are the Cartesian coordinates of \ce{N}, \ce{S} and \ce{F} atoms, respectively. Another relevant indicator is, CN$_i$: the numbers of coordinated F atoms around Al/Ca, and it is also calculated for each configuration $i$. Therefore, an alternative characterization is $\mathbf{G'} = \{ [\langle \mathbf{g'}_i \rangle, \langle \text{CN}_i \rangle ]_i \}$, in which state $9$, the state previous to any product state, is further clustered into $6$ sub-states. In this way, we gain more detailed information about the reaction mechanism.

\section{Results and discussion}
Metal-fluorine diatomic molecules are used for many precision spectroscopy and laser cooling applications. Such molecules can be produced by a chemical reaction between hot ablated metal atoms and a fluorine-containing gas. Here, we analyze the two most-commonly used fluorine-donor reaction gases, \ce{NF3} and \ce{SF6} and compare the reaction efficiencies to form \ce{AlF} and \ce{CaF}.

\begin{figure}[h]
    \centering
    \includegraphics[width=0.8\linewidth]{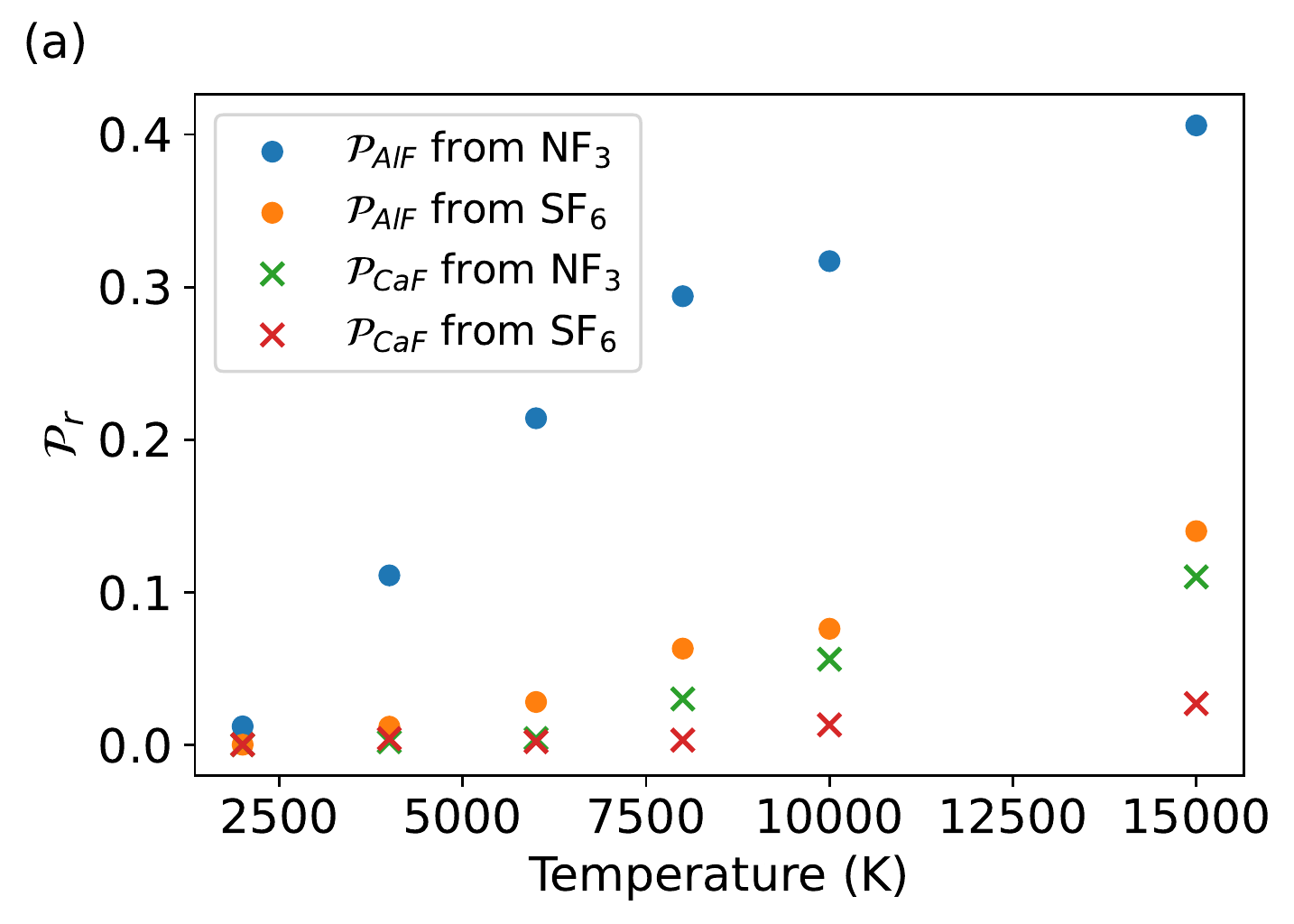}\\
    \includegraphics[width=0.8\linewidth]{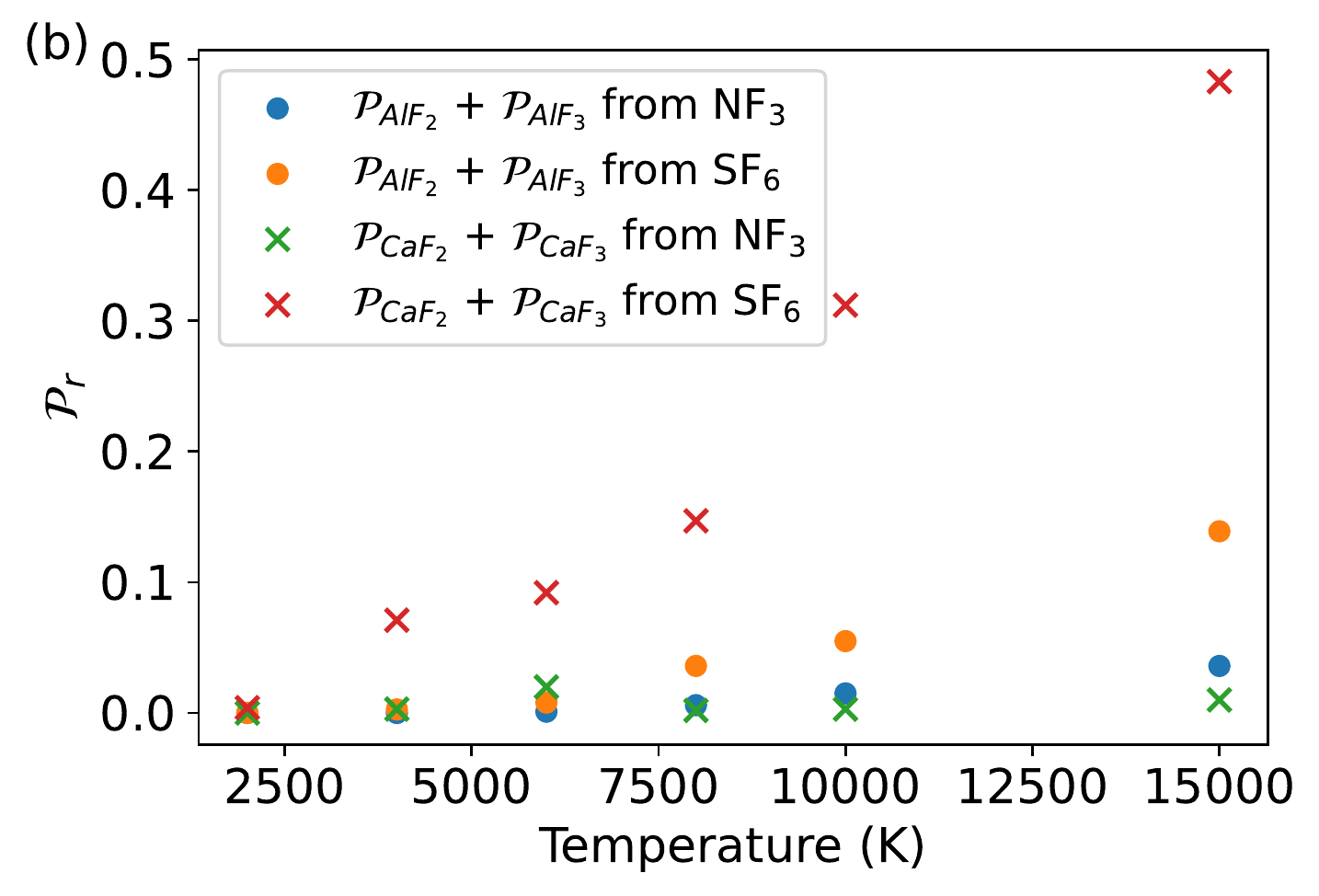}
    \caption{Reaction probability of (a) AlF/CaF and (b) AlF$_n$/CaF$_n$ by-products for hot collisions of Al/Ca with \ce{SF6} and \ce{NF3} gases as a function of the temperature.}
    \label{Fig:productivity_AlF_CaF_MgF}
\end{figure}

\begin{figure*}[t!!!]
    \centering
    \includegraphics[width=1\linewidth]{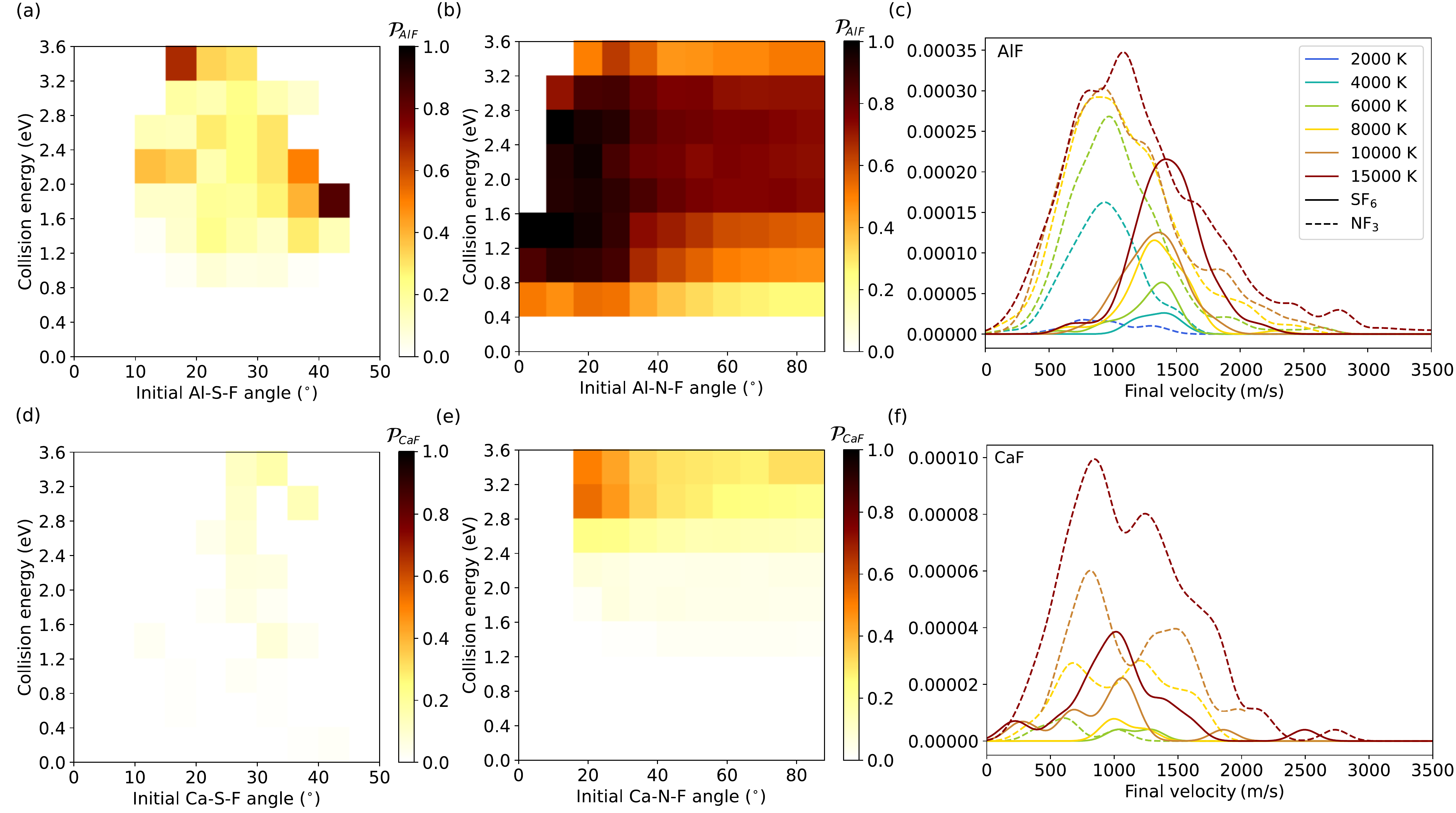}
    \caption{(a) Reaction probability to produce AlF by using \ce{SF6}. (b) Reaction probability of AlF using \ce{NF3}. (c) Velocity distribution of AlF, normalized to the corresponding reaction probabilities at different temperatures. (d) Reaction probability of CaF using \ce{SF6}.(e) Reaction probability of CaF using \ce{NF3}. (f) Velocity distribution of CaF, normalized to the corresponding reaction probabilities at different temperatures.}
    \label{Fig:productivity_vs_collision_SF6_NF3_Al_Ca}
\end{figure*}

\subsection{Reaction probability}\label{subsection:productivity}
The formation of AlF and CaF molecules in Al/Ca + \ce{SF6}/\ce{NF3} collisions have been studied, and the results are shown in Fig.~\ref{Fig:productivity_AlF_CaF_MgF}. This figure displays the reaction probability for different products as a function of the temperature. At all temperatures considered, AlF is produced more efficiently than CaF irrespective of the type of reactant gas used in the reaction. Indeed, this behavior has been experimentally observed when comparing the brightness of CaF and AlF beams emerging from a buffer gas cell~\cite{Buffergas2, Hofsaess2021}. However, in the case of by-products, as shown in panel (b) of Fig.~\ref{Fig:productivity_AlF_CaF_MgF},  \ce{CaF2} is produced more efficiently than \ce{AlF2} and \ce{AlF3} in both gases. At higher temperatures, the reaction probability of producing AlF and CaF is more prominent than at lower ones, as expected in the case of reactions with a barrier. However, for the entire range of temperatures under consideration, the reaction probability for AlF and CaF via \ce{NF3} is higher than in the case of \ce{SF6}, as shown in panel (b) of Fig.~\ref{Fig:productivity_AlF_CaF_MgF}. In particular, in the case of AlF formation, this difference is as large as an order of magnitude for particular temperatures. Therefore, Al/Ca + \ce{NF3} $\rightarrow$NF$_2$ + AlF/CaF reactions have a lower reaction barrier than Al/Ca + \ce{SF6} $\rightarrow$SF$_5$ + AlF/CaF and Al/Ca reactions. Indeed, the activation energy for Al + \ce{SF6} and Al + \ce{NF3} reactions have been experimentally determined and they have a value of 9.5~kcal/mol (4781~K) and 5.99~kcal/mol (2990~K)~\cite{parker2002kinetics}, respectively. Consequently, the use of \ce{NF3} in the buffer gas cell should lead to a brighter beam than using \ce{SF6}. Moreover, using \ce{NF3} reduces the number of by-products and possible contamination of the buffer gas cell.

One reason for the higher reaction probability to produce AlF or CaF molecules when using \ce{NF3} is due to the difference in bond energy of F-atoms in \ce{NF3} and \ce{SF6}. For \ce{NF3} the bond energy is (2.9~eV)~\cite{NF3}, which is 1.1~eV lower than the bond energy of F-atoms in \ce{SF6} ~\cite{SF6} (4.0~eV). Therefore, it is easier to remove a fluorine atom from \ce{NF3} than from \ce{SF6}, leading to a higher reaction probability at a given temperature. In addition, the probability of formation of AlF molecules is higher than CaF because the bond energy of AlF (6.9~eV) is 1.4~eV larger than that of CaF (5.5~eV). This simple picture is substantiated in Section~\ref{subsection:stereochemsitry} which discusses the role of stereochemistry.

\begin{figure*}[t]
    \centering
    \includegraphics[width=1\linewidth]{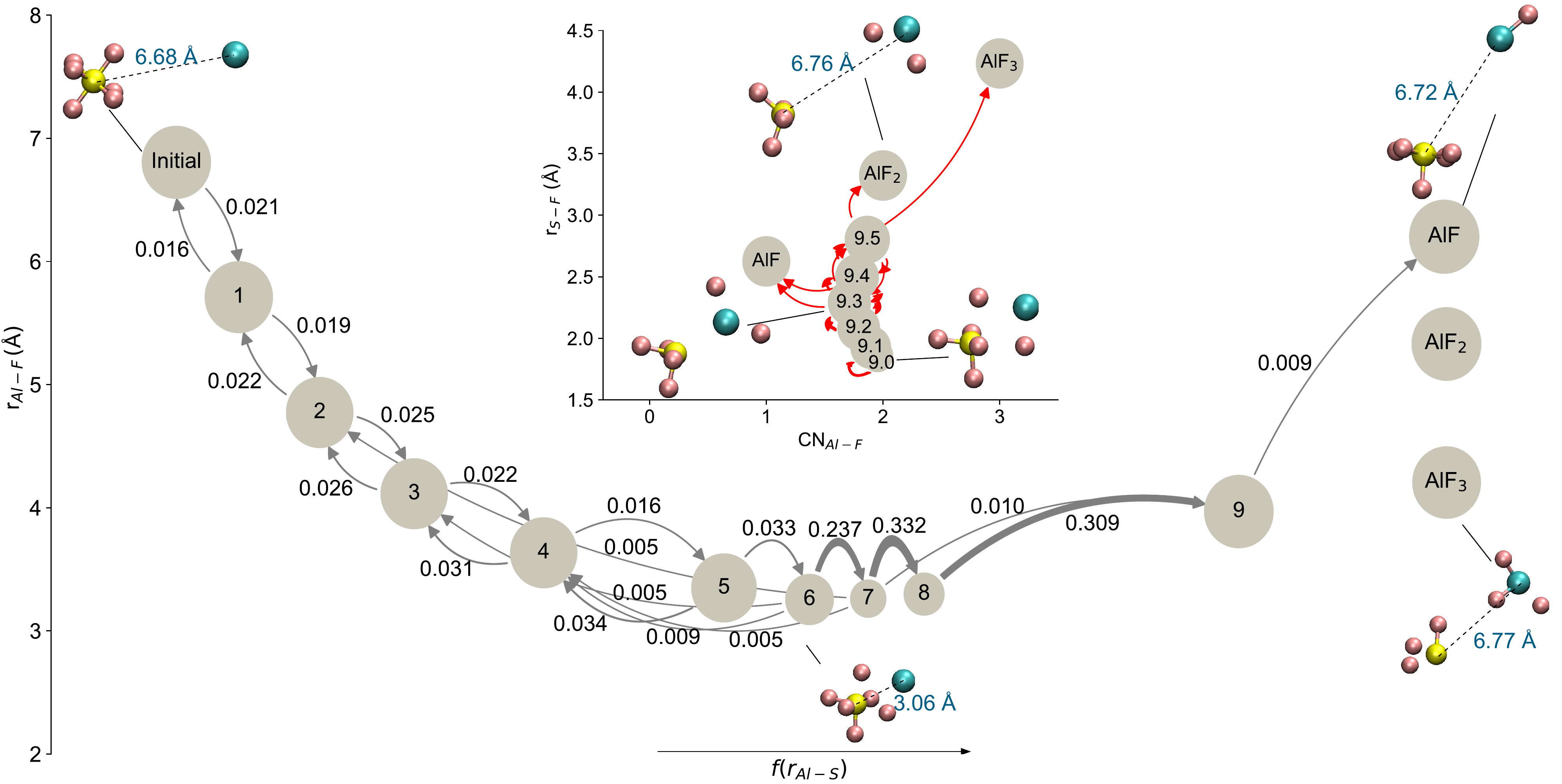}
    \caption{Reaction model of the \ce{SF6}~+~Al reaction at 10000~K. The nodes are the clustered states based on the structural similarity of the sampled configurations, labeled as 0 (initial state) to 12 (products). The arrows indicate the transition probability between two states $p_{ij}$. The model is projected to the $ (f(r_{\text{Al-S}}), r_{\text{Al-F}} )$ space, where $r_{\text{Al-S}}$,$ r_{\text{Al-F}}$ are the average Al-S, Al-F distances of each state, respectively. The inset shows the detailed evolution from state $9$ to different products, represented in the (CN$_{\text{Al-F}}, r_{\text{S-F}}$) space, where CN$_{\text{Al-F}}$ is the average number of coordinated F atoms around Al atom of the states. The S, F, and Al atoms are shown in yellow, coral, and cyan, respectively. The blue numbers are the average $r_{\text{Al-S}}$ of the states. To show the detailed transitions in the product region, the threshold showing the arrows (in red) in the inset panel is smaller than the main figure.}
    \label{Fig:reaction_model_SF6_Al}
\end{figure*}

A more detailed treatment of the temperature dependence is presented in panels (a-b) and (d-e) of Fig.~\ref{Fig:productivity_vs_collision_SF6_NF3_Al_Ca}. Here, we display the reaction probability as a function of collision energy and initial angle of the metal atom, as introduced in Fig.\ref{Fig:initial_positions}. In the case of \ce{SF6}, the reaction producing either AlF or by-products generally occurs at relatively high collision energies ($\gtrsim 0.8$~eV$\sim$9300~K). On the contrary, AlF is produced at much lower collision energies of $\gtrsim 0.4$~eV~$\sim$4600~K when \ce{NF3} is used. Reaction by-products appear only at very high energies $\gtrsim 1.6$~eV ($\sim$18600~K) in both \ce{SF6} and \ce{NF3}. This suggests that such reactions require the activation of F atoms promoted by the collision with an Al atom. However, the production of by-products requires more F atoms to be activated. This is in stark contrast to reactions that involve Ca atoms, for which the production of CaF or \ce{CaF2} requires a similar collision energy ($\gtrsim 0.8$~eV) when reacting with \ce{SF6}. Moreover, this reaction preferentially produces \ce{CaF2} and not CaF. Unlike Al, in the reaction of Ca + \ce{NF3}, CaF is produced also at high collision energies $\gtrsim 1.2$~eV ($\sim$13900~K), while the production of \ce{CaF2} occurs at a wider range of collision energies. These phenomena suggest that Al and Ca experience very different reaction mechanisms when reacting with \ce{SF6} and \ce{NF3}.

\subsection{Stereochemistry}\label{subsection:stereochemsitry}

The orientation of the reactants affects the reactivity of a given chemical reaction. These effects are due to subtleties in the underlying energy landscape of every molecular interaction, such as geometry effects or local equilibrium states. Here, by looking into the reaction probability as a function of the atom's angle of incidence, we can evaluate selectivity effects based on the orientation of the interacting partners, as presented in panels (a), (b), (d), and (e) of Fig.~\ref{Fig:productivity_vs_collision_SF6_NF3_Al_Ca}. Concretely, it informs us about the anisotropy of the interaction and the geometry effects on the interaction energy. 

First, one notices that the production of AlF/CaF via \ce{NF3} occurs over a wider range of angles than in the case of \ce{SF6}, indicating a more isotropic interaction. Similarly, when comparing the production of CaF and AlF in \ce{SF6}, we notice that CaF is formed only at incident angles close to 30$^{\circ}$, whereas AlF is formed over a range of angles between 10$^{\circ}$ and 45$^{\circ}$. The same effect, although not as pronounced, is observed in the case of \ce{NF3}. AlF is produced at angles between 0 and 90$^{\circ}$, and CaF for angles between 20$^{\circ}$ and 90$^{\circ}$. Therefore, AlF formation is less selective than CaF, which may impact the reaction probability.   

\begin{figure*}[t]
    \centering
    \includegraphics[width=1\linewidth]{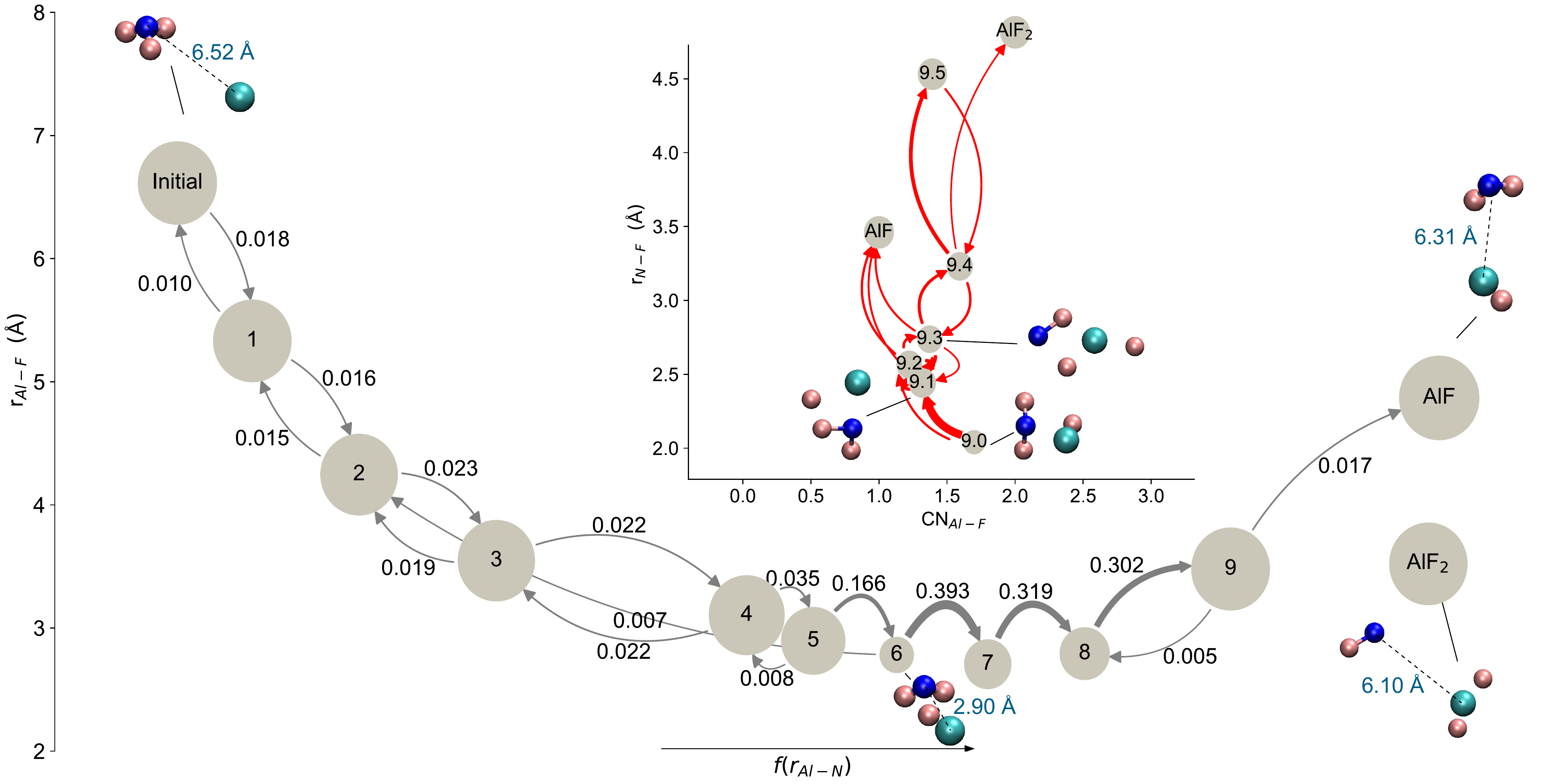}
    \caption{Reaction model of the \ce{NF3}+Al reaction at 10000K. The nodes are the clustered states based on the structural similarity of the sampled configurations, labeled as 0 (initial state) to 11 (products). The arrows indicate the transition probability between two states $p_{ij}$. The model is projected to the $ (f(r_{\text{Al-N}}), r_{\text{Al-F}} )$ space, where $r_{\text{Al-N}}$,$ r_{\text{Al-F}}$ are the average Al-N, Al-F distances of each state, respectively. The inset shows the detailed evolution from state $9$ to different products, represented in the (CN$_{\text{Al-F}}, r_{\text{N-F}}$) space, where CN$_{\text{Al-F}}$ is the average number of coordinated F atoms around Al atom of the states. The N, F, and Al atoms are shown in blue, coral, and cyan, respectively. The blue numbers are the average $r_{\text{Al-N}}$ of the states. To show the detailed transitions in the product region, the threshold showing the arrows (in red) in the inset panel is smaller than the main figure.}
    \label{Fig:reaction_model_NF3_Al}
\end{figure*}

\subsection{Velocity distribution of the products}

The generation of AlF and CaF with either reactant gas releases several thousand kelvins of energy, comparable to or larger than the collision energy. This must be carried away either in translational motion of the products or by their internal (electronic, vibrational, and rotational) energy and subsequently cooled by the buffer gas. The results for the velocity distribution of AlF and CaF are shown in panels (c) and (f) of Fig.~\ref{Fig:productivity_vs_collision_SF6_NF3_Al_Ca}. In particular, we notice that \ce{SF6} leads to a narrower velocity distribution than \ce{NF3} independent of the colliding atom since the bond energy of F-atoms in  \ce{SF6} is 1.1~eV larger than in \ce{NF3}, leading to a lower exothermicity. In addition, we notice that \ce{NF3} as a reactant leads to a more significant number of molecules at low velocities. This behavior may be correlated with the stereochemistry of the reaction: low incident angles lead to a significant reaction probability in Al/Ca + \ce{NF3} reactions, whereas Al/Ca +\ce{SF6} reactions show the most considerable reaction probability at larger incident angles. Indeed, a lower incident angle means a more efficient energy transfer between the metal atom and the F atom as opposed to the case of large incident angles, in which different F atoms and internal excitation of the F-containing molecule can play a role in the dynamics. Therefore, it is preferable to use \ce{NF3} to obtain colder beams from a buffer gas cell. Finally, we realize that the temperature has only a subtle influence on the most probable velocity (or distribution mode). This is expected, since the temperature has only a minimal impact on the reaction channels.

\subsection{By-products}
\begin{figure*}[t]
    \centering
    \includegraphics[width=1\linewidth]{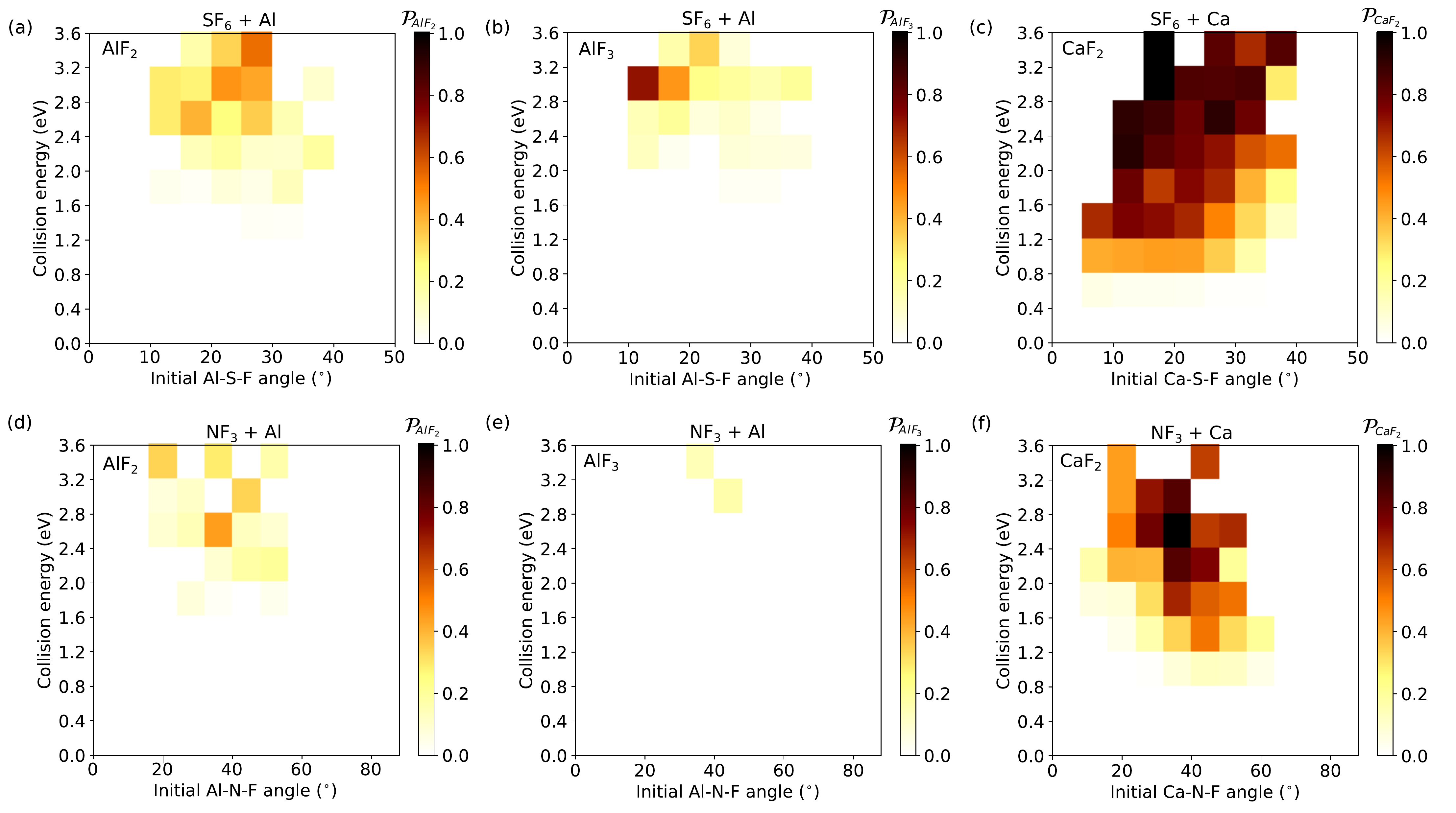}
    \caption{Reaction probability producing different by-products as a function of initial Al-S-F angle and collision energy in the \ce{SF6}/\ce{NF3} + Al/Ca reactions. (a) Reaction probability of \ce{AlF2} via \ce{SF6} + Al. (b) Reaction probability of \ce{AlF3} via \ce{SF6} + Al. (c) Reaction probability of \ce{CaF2} via \ce{SF6} + Ca. (d) Reaction probability of \ce{AlF2} via \ce{NF3} + Al. (e) Reaction probability of \ce{AlF3} via \ce{NF3} + Al. (f) Reaction probability of \ce{CaF2} via \ce{NF3} + Ca.}
    \label{Fig:productivity_vs_collision_SF6_AlF23_CaF2}
\end{figure*}

In this section, we explore the production of by-products in Al + \ce{SF6}/\ce{NF3} and Ca + \ce{SF6}/\ce{NF3} reactions. The former reaction may produce AlF$_2$ and AlF$_3$ molecules as by-products, whereas the latter leads to CaF$_2$. The results are summarized in Fig.~\ref{Fig:productivity_vs_collision_SF6_AlF23_CaF2}. In this figure, we notice that the reaction probabilities for the formation of by-products behave differently for AlF and CaF. For instance, reactions with \ce{SF6} show a higher reaction probability to form by-products than in the case of \ce{NF3}, which may be related to the fact that \ce{SF6} provides more F atoms than \ce{NF3}. Similarly, we observe that Ca reactions lead to a higher probability of by-products than reactions involving Al. Indeed, as discussed in section \ref{subsection:reactionmodel}, the transition probability to \ce{CaF2} is higher than \ce{CaF}. Therefore, \ce{CaF2} is the main product of the reactions involving Ca.

Reactions with Al show a lower threshold energy for the formation of by-products than in the case of Ca reactions, as it is shown in Fig.~\ref{Fig:productivity_vs_collision_SF6_AlF23_CaF2}. As expected, the collision energy required to form \ce{AlF2} is higher than for the formation of AlF but lower than for \ce{AlF3}.  On the contrary, the collision energy required to form \ce{CaF2} is lower than \ce{CaF}. As discussed in Sec.~\ref{subsection:reactionmodel}, \ce{CaF} is mainly formed through the dissociation of \ce{CaF2}, which can be promoted when the kinetic energy of \ce{CaF2} increases. Similar phenomena can also be observed in the reactions involving \ce{NF3}, as shown in Figs.~\ref{Fig:productivity_vs_collision_SF6_NF3_Al_Ca} and \ref{Fig:productivity_vs_collision_SF6_AlF23_CaF2}.


\subsection{Tree-shaped reaction model}\label{subsection:reactionmodel}

\begin{figure*}[t]
    \centering
    \includegraphics[width=1\linewidth]{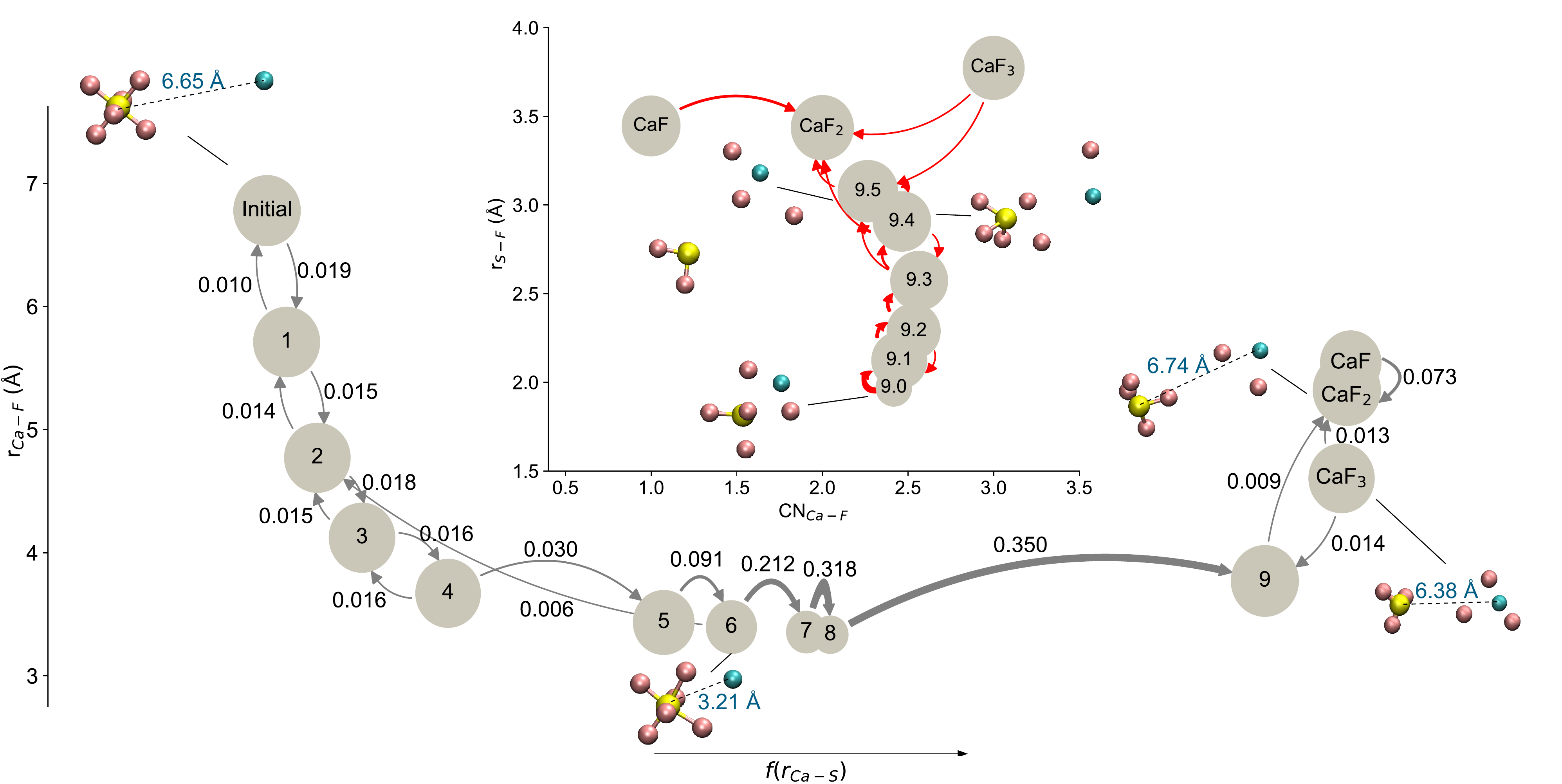}
    \caption{Reaction model of the \ce{SF6}+Ca reaction at 10000~K. The nodes are the clustered states based on the structural similarity of the sampled configurations, labeled as 0 (initial state) to 11 (products). The model is projected to the $ (f(r_{\text{Ca-S}}), r_{\text{Ca-F}} )$ space, where $r_{\text{Ca-S}}$,$ r_{\text{Ca-F}}$ are the average Ca-S, Ca-F distances of each state, respectively. The arrows indicate the transition probability between two  states $p_{ij}$. The inset shows the detailed evolution from state $9$ to different products, represented in the (CN$_{\text{Ca-F}}, r_{\text{S-F}}$) space, where CN$_{\text{Ca-F}}$ is the average number of coordinated F atoms around Ca atom of the states. The N, F, and Ca atoms are shown in blue, coral, and cyan, respectively. The blue numbers are the average $r_{\text{Ca-S}}$ of the states. To show the detailed transitions in the product region, the threshold showing the arrows (in red) in the inset panel is smaller than the main figure.}
    \label{Fig:reaction_model_SF6_Ca}
\end{figure*}

\begin{figure*}[t]
    \centering
    \includegraphics[width=1\linewidth]{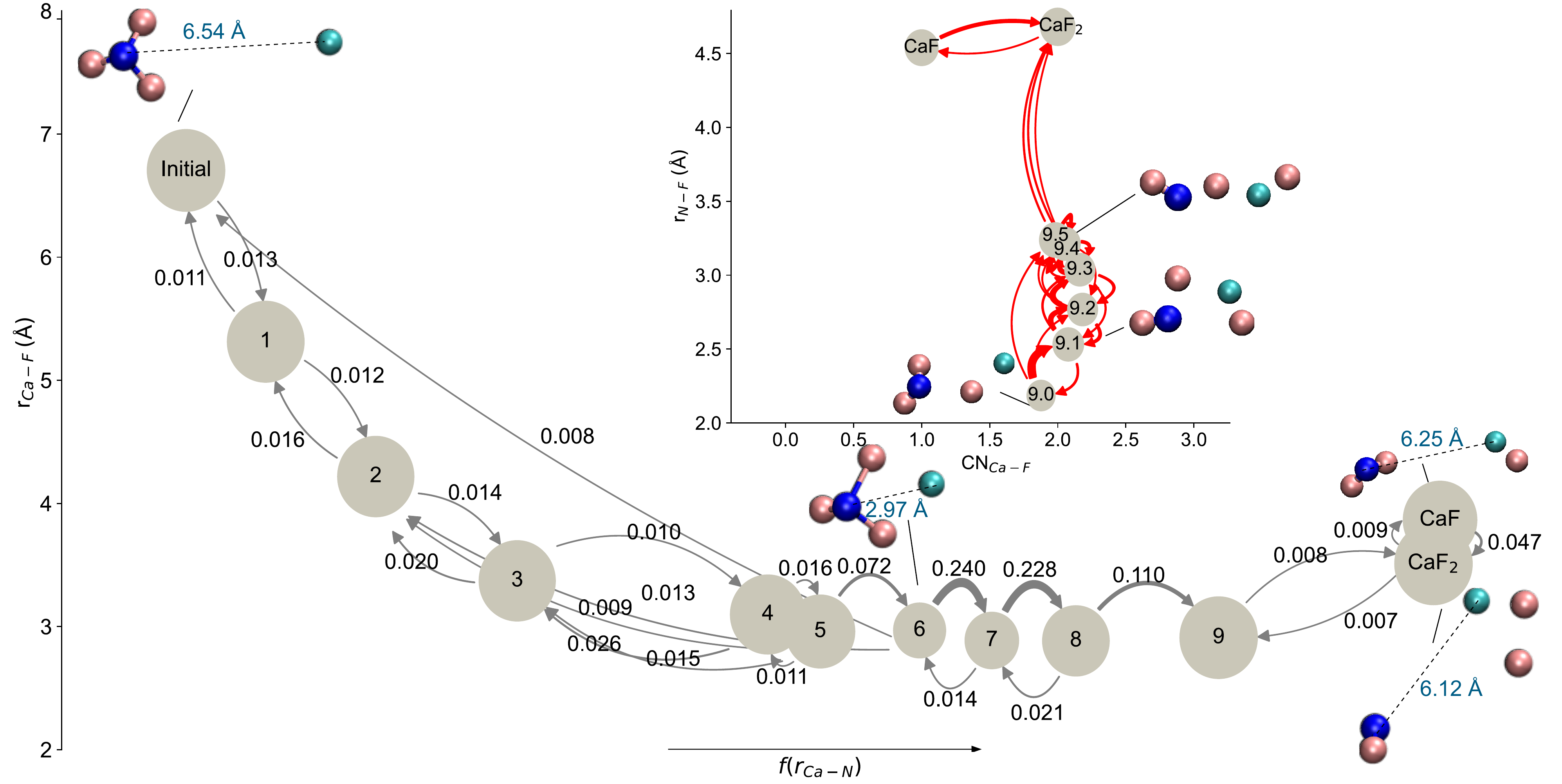}
    \caption{Reaction model of the \ce{NF3}+Ca reaction at 10000~K. The nodes are the clustered states based on the structural similarity of the sampled configurations, labeled as 0 (initial state) to 11 (products). The model is projected to the $ (f(r_{\text{Ca-N}}), r_{\text{Ca-F}} )$ space, where $r_{\text{Ca-N}}$,$ r_{\text{Ca-F}}$ are the average Ca-N, Ca-F distances of each state, respectively. The arrows indicate the transition probability between two  states $p_{ij}$. The inset shows the detailed evolution from state $9$ to different products, represented in the (CN$_{\text{Ca-F}}, r_{\text{N-F}}$) space, where CN$_{\text{Ca-F}}$ is the average number of coordinated F atoms around Ca atom of the states. The N, F, and Ca atoms are shown in blue, coral, and cyan, respectively. The blue numbers are the average $r_{\text{Ca-N}}$ of the states. To show the detailed transitions in the product region, the threshold showing the arrows (in red) in the inset panel is smaller than the main figure.}
    \label{Fig:reaction_model_NF3_Ca}
\end{figure*}

So far, we have identified that the collision energy and angle of incidence play a relevant role in the reaction probability to produce AlF/CaF and by-products via \ce{SF6} and \ce{NF3}. This section presents models for these reactions which help to elucidate the underlying reaction mechanisms hidden in the AIMD simulations. These models are presented for a temperature of 10000~K, although they are readily extensible to any temperature.

A chemical reaction can be modeled by looking into different reaction channels weighted by corresponding transition probabilities. In particular, every chemical reaction under consideration is modeled by a tree-shaped model using 13 states. Consequently, the backward transitions should be considered independent of the forward transitions. The results for \ce{Al} + \ce{SF6} collisions are shown in Fig.~\ref{Fig:reaction_model_SF6_Al}. This reaction is sequential since only one major forward transition appears. In other words, no significant competition exists between transition branches in the product nodes. All states apart from state 8 show a backward and forward transition probability. As a result, backward transitions are only observed before the system evolves into state $8$: these trajectories constitute elastic events. Provided the system reaches state $8$, it will experience either a reactive or inelastic scattering process. The relative long distance between state $8$ and state $9$ indicates that the products \ce{AlF_{x}} ($x$=2,3) have been separated from target molecules, and the state $9$ will continue to evolve, producing \ce{AlF} and by-products: \ce{AlF2} or \ce{AlF3}.

In the Al+\ce{SF6} reaction, \ce{S-F} distances of the intermediate states correlates with the activity of \ce{F} atoms, as shown in the inset of Fig.~\ref{Fig:reaction_model_SF6_Al}, where the state $9$ has been further clustered into 6 sub-clusters with increasing $r_{\ce{S-F}}$. From state $9.0$ to state $9.5$, $r_{\ce{S-F}}$ becomes larger, suggesting that more \ce{F} atoms are activated. Furthermore, there are significant transitions between states from $9.2$ to $9.5$, which can be considered as a resonance between the \ce{SF6} molecule with fewer or more activated \ce{F} atoms during the generation of products from the target molecule. Transitions between the products and sub-clusters of state $9$ show that \ce{AlF} is produced from state $9.3$ and $9.4$, while by-products, \ce{AlF2} and \ce{AlF3}, are produced from state $9.5$. Comparing $r_{\ce{S-F}}$ of state $9.3$ and $9.4$ with the one of state $9.5$, it can be concluded that \ce{AlF} is produced from the state with fewer \ce{F} atoms activated, while \ce{AlF2} and \ce{AlF3} are produced from states with more \ce{F} atoms activated. Therefore, the activation of \ce{F} atoms in \ce{SF6} is crucial for the selectivity of the reaction, which is in agreement with the reaction probability-collision energy relationships shown in Fig.~\ref{Fig:productivity_vs_collision_SF6_AlF23_CaF2} (a)-(b). As discussed in section \ref{subsection:productivity}, \ce{AlF} is produced more efficiently at relatively low collision energies compared with \ce{AlF2} and \ce{AlF3}. Indeed, lower collision energy will activate relative fewer \ce{F} atoms and therefore produce more \ce{AlF}. On the contrary, higher collision energies will activate more \ce{F} atoms, thus, producing \ce{AlF2} or \ce{AlF3}.

Interestingly enough, the above discussion regarding Al +\ce{SF6} reactions is applicable to the Al + \ce{NF3}  reaction, as shown in Fig.~\ref{Fig:reaction_model_NF3_Al}, but it is unsuited to reactions involving \ce{Ca}. Fig.~\ref{Fig:reaction_model_SF6_Ca} shows a tree-shaped reaction model for Ca + \ce{SF6}, which is similar to Al + \ce{SF6}. However, unlike Al + \ce{SF6}, Ca + \ce{SF6} collisions end up producing \ce{CaF2}, showing the highest reaction probability, as shown in section~\ref{subsection:stereochemsitry}, over by-products. Indeed, \ce{CaF} and \ce{CaF3} are produced from further reaction of \ce{CaF2}, which is supported by significant transitions between \ce{CaF2}, \ce{CaF} and \ce{CaF3} shown in the inset of Fig.~\ref{Fig:reaction_model_SF6_Ca}. The preference towards producing \ce{CaF2} is also observed in \ce{Ca}  + \ce{NF3} reaction shown in Fig.~\ref{Fig:reaction_model_NF3_Ca}. Therefore, \ce{Ca} atom strongly prefers to establishes a chemical bond with two \ce{F} atoms rather than with a single one, which can be related with the fact that CaF is a radical as opposed to a stable molecule as AlF. Further interactions between \ce{CaF2} and \ce{SF4}/\ce{NF} will make \ce{CaF2} lose or capture one \ce{F} atom. Finally, our reaction model seems to indicate that reactions involving Ca atoms will lead to a larger number of by-products and a larger degree of contamination of the buffer gas cell. 

\section{Conclusions and outlook}

\begin{figure}[h]
    \centering
    \includegraphics[width=1\linewidth]{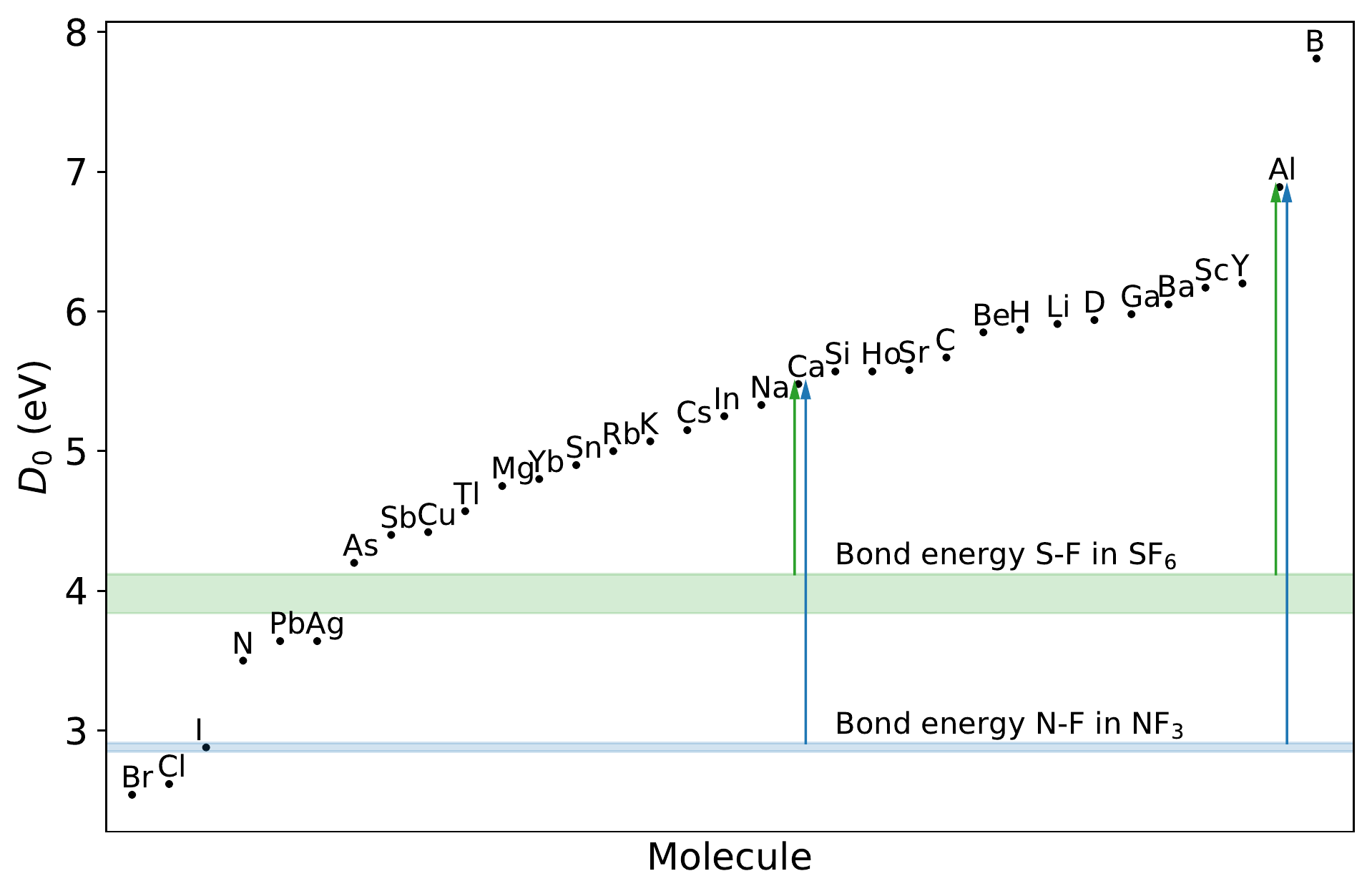}
    \caption{The dissociation energy of diatomic monofluoride molecules XF, for different atoms X, taken from Ref.~\cite{database}. The two red lines denote the limits for the bond energy of S-F in \ce{SF6} whereas the black lines denote the same magnitude for \ce{NF3}. The arrows indicate the exothermic energy of the reactions to form CaF and AlF. }
    \label{Fig:XF}
\end{figure}

In this work, we have shown that the reaction probability of forming AlF and CaF via ablation of metal atoms in an F-containing reactant gas is higher when using \ce{NF3} as a reactant gas than in the case of \ce{SF6}. Indeed, this effect seems to relate to the reaction's exothermicity: the more significant the difference between the binding energy of the product (AlF/CaF) and the bond energy of fluorine into the reactant molecule, the larger the reaction probability is. In particular, the exothermicity for AlF from \ce{NF3}, AlF from \ce{SF6}, CaF from \ce{NF3} and CaF from \ce{SF6} is given by 4.0~eV, 2.9~eV, 2.6~eV and 1.5~eV, respectively, in line with the observed probabilities shown in panel (a) of Fig.~\ref{Fig:productivity_AlF_CaF_MgF}. These are further depicted in Fig.~\ref{Fig:XF}, in which the exothermic energy corresponds to the length of the arrows. Therefore, it is easier for a hot metal atom to take a fluorine atom from an \ce{NF3} than from an \ce{SF6} molecule. Moreover, Fig.~\ref{Fig:XF} shows a detailed list of XF molecules, taken from Ref.~\cite{database} that can be formed via exothermic processes utilizing \ce{NF3} and \ce{SF6} gases. As a result, and concerning the lower binding energy of the N-F bond in \ce{NF3}, many XF molecules could be formed in a buffer gas source. In other words, \ce{NF3}, as the fluorine-donor gas, will help to explore more fluorine-containing diatomic molecules. Given this, \ce{XeF2} is anticipated to be a good candidate as an F-atom donor based on its very low bond energy in comparison with AlF and CaF molecules. 


We have shown that different fluorine-donor molecules in a buffer gas cell environment have an important impact on the target molecules' production efficiency and velocity distribution. In particular, we have demonstrated that the Ca/Al + \ce{NF3} reaction is a more efficient route towards the production of CaF and AlF molecules than the Ca/Al + \ce{SF6} one. Indeed, the difference in reaction probability can be as large as one order of magnitude. In addition, we have identified the main reaction mechanisms for those reactions using a tree-shaped reaction model. Our results indicate that the buffer gas cell's amount of by-products and possible contamination depends on the fluorine-donor molecule. The velocity distribution of the products depends drastically on the reactants. For instance, we notice that Ca/Al + \ce{NF3} lead to a broader velocity distribution than Ca/Al + \ce{SF6}. The higher the reaction efficiency with \ce{NF3} means that a significantly lower flow rate of reactants gas can be used. This reduces contamination and the build-up of ice in the cell, preventing efficient thermalization of the molecules with the cryogenic helium. In addition, the lower velocity of the reactants reduced the number of collisions required to cool the molecules. A lower helium flow reduces the overall gasload in the system and allows for a faster extraction from the cell, which is highly advantageous for experiments that are sensitive to collisions with helium or benefit from short molecular pulses.

Finally, we can conclude that, in general, it would be better to use \ce{NF3} as a fluorine-donor molecule than \ce{SF6} for forming metal-fluorine molecules. It is worth emphasizing that a better understanding of the chemistry in a buffer gas helps design better buffer gas cells towards achieving brighter and colder molecular beams, which are necessary to exploit all the potential of ultracold molecules. Therefore, we hope our work motivates the community to explore more possibilities and new ways to obtain molecules in buffer gas sources.

\section{Acknowledgments}

X. Liu and J. P.-R. thanks the support of the Deutsche Forschungsgemeinschaft (DFG – German Research Foundation) under the grant number PE 3477/2 - 493725479.

\section{Data Availability Statement}
The data that support the findings of this study are available from the corresponding author upon reasonable request.

\bibliography{buffergas}
\end{document}